%% file: main.tex
\documentclass[
 reprint,
 amsmath,amssymb,
 aps,
 prd,
]{revtex4-2}

\usepackage{graphicx}
\usepackage{dcolumn}
\usepackage{bm}
\usepackage{multirow}
\usepackage[table]{xcolor}
\usepackage{epsfig}
\usepackage{url}
\usepackage{aas_macros}

\usepackage[normalem]{ulem}

\usepackage{latexsym}
\usepackage{epsfig}
\usepackage{amsmath}
\usepackage{amssymb}
\usepackage{wasysym}
\usepackage{graphicx}
\usepackage{verbatim}
\usepackage{enumerate,mdwlist}
\usepackage[titletoc]{appendix}
\usepackage{amsfonts}
\usepackage{pgfplots}
\usepackage[export]{adjustbox}
\usepackage{bbold}

\usepackage[linktocpage=true]{hyperref}
\hypersetup{
colorlinks=true,
citecolor=indigo(dye),
linkcolor=indigo(dye),
urlcolor=indigo(dye),
pdfauthor={},
pdftitle={},
pdfsubject={}
}

\input{new_commands}

\begin{document}

\preprint{APS/123-QED}

\title{Effects of Galaxy Cluster Structure on Lensed Gravitational Waves}

\author{Luka Vujeva}
\email{luka.vujeva@nbi.ku.dk}
\author{Jose Mar\'ia Ezquiaga}
\author{Rico K.~L.~Lo}
\author{J. C.~L.~Chan}
\affiliation{Center of Gravity, Niels Bohr Institute, Blegdamsvej 17, 2100 Copenhagen, Denmark}

\date{\today}

\begin{abstract}
Strong gravitational lenses come in many forms, but are typically divided into two populations: galaxies, and groups and clusters of galaxies. 
The largest objects in the Universe (i.e. galaxy clusters) are highly irregular and composed of many components due to a history of (or active) hierarchical mergers. In this work, we analyze the discrepancies in the observables of strongly lensed gravitational wave transients in both scenarios, 
namely relative magnifications, time delays, and image multiplicities.  
We compare the detection rates 
between the single spherical dark matter halo models found in the literature, and publicly available state-of-the-art cluster lens models.
We find there to be approximately an order of magnitude fewer detection of strongly lensed transients in the realistic model case, 
likely caused by their loss of overall strong lensing optical depth. 
We also report detection rates
in the weak lensing or single-image regime.
Additionally, we find a systemic shift towards lower time delays between the brightest image pairs in the cases of the realistic models, as well as higher fractions of positive versus negative parity images, which was previously reported in the literature.
This deviation in the joint relative magnification factor-time delay distribution will hinder the feasibility of the reconstruction of cluster-scale lenses through gravitational wave transients alone, but can still provide a lower limit on the lens mass.
\end{abstract}

\maketitle


\section{\label{sec:intro}Introduction}

\begin{figure*}[t]
    \centering
    \includegraphics[width=\linewidth]{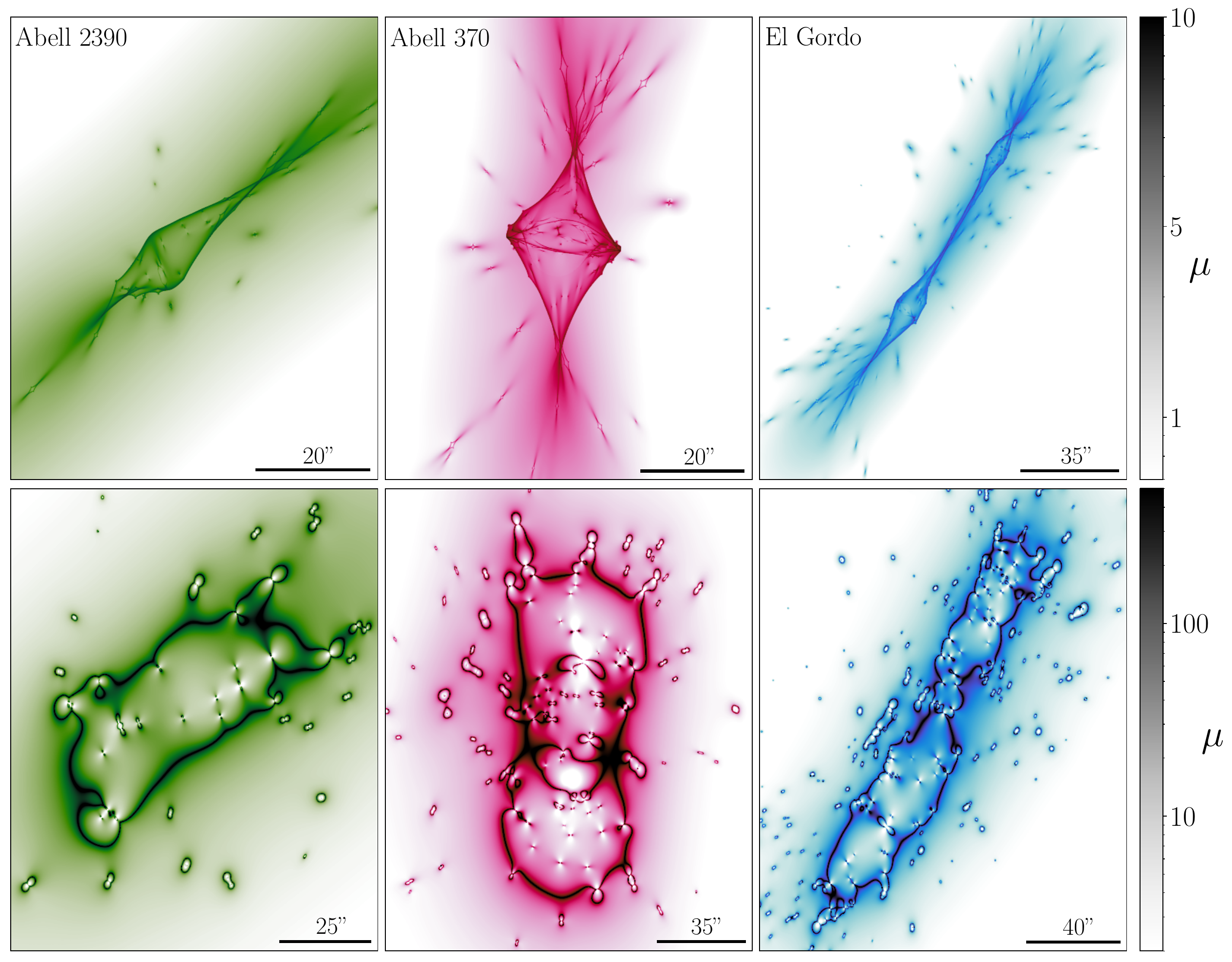}
    \caption{Source plane (top) and image plane (bottom) absolute magnification ($\mu$) maps of Abell 2390 (left, green), Abell 370 (middle, pink) and El Gordo (right, blue) for a source at $z=3$. These three clusters exhibit a rich morphology. They have similar total mass, $\sim10^{15}M_\odot$, but different number of member galaxies. They are also located at different redshifts, increasing from left to right (see Table \ref{tab:clusters} for details). This changes the angular size, which can be compared to the reference scale in the solid black line.
    }
    \label{fig:muclust}
\end{figure*}

Gravitational lensing is an invaluable tool that allows us to learn about dark matter \cite{2010CQGra..27w3001B}, as well as providing a glimpse into the high-redshift Universe.
Although the overwhelming majority of observed strongly lensed sources we observe are static sources (i.e. galaxies, quasars, or stars), we have begun to observe lensed transients, such as supernovae \cite{kelly2015sn,goobar2017sn,Rodney:2021keu, chen2022sn, kelly2022sn,goobar2023sn,frye2024sn,pierel2024sn}. 
Despite the low number of observed lensed transients, this number is expected to increase dramatically in the advent of new telescopes such as Vera Rubin Observatory \cite{lsst2019} and Euclid \cite{euclid2022} for supernovae, and future gravitational wave (GW) detectors \cite{LIGOScientific:2016wof,2020JCAP...03..050M}.
Although we have detected strongly lensed transients in the electromagnetic (EM) spectrum, a strong lensing multi-messenger event remains elusive. Even one detection would provide a stringent tests of general relativity, cosmology, and the structure of dark matter halos \cite{2024SSRv..220...12S,Oguri:2019fix,Ezquiaga:2020dao,Tambalo:2022wlm,Caliskan:2023zqm, Savastano:2023spl, Brando:2024inp, Chan:2025wgz}. However, given the rarity of strong lensing \cite{2020MNRAS.495.3727R}, along with current LIGO \cite{LIGOScientific:2014pky}, VIRGO \cite{VIRGO:2014yos}, and KAGRA \cite{KAGRA:2020tym} GW detector sensitivities,
we have still yet to identify a strongly lensed GW \cite{Hannuksela:2019kle,LIGOScientific:2021izm,Janquart:2023mvf,LIGOScientific:2023bwz}.

The expected optical depth ($\tau$) has been estimated to be $\tau \sim 10^{-3}$ for sources up until $z\sim1$, see e.g. \cite{Oguri:2019fix,Xu:2021bfn}.
These studies assume single dark matter halos, which is reasonable for galaxy-scale lenses (which these studies mainly consider).
However, this assumption begins to break down as we ascend in total dark matter halo mass ($M_\mathrm{halo} \sim 10^{13}-10^{15} M_\odot$), where we must begin to consider the structure of galaxy groups, and in the case of this work, galaxy clusters. 

Galaxy clusters have been extensively studied due to their incredible lensing efficiency, and their prowess as gravitational telescopes, which allow for detection of galaxies \cite{2023ApJ...957L..34W,2024arXiv241113640K} and stars \cite{Welch_2022, chema2023} at redshifts beyond the detection horizon of current telescopes \cite{mason2015,vujeva2024}. A key result from the studies of dark matter content of galaxy clusters is that they are complex structures comprised of many dark matter halos \cite{eliasdottir2007, furtak2023, gledhill2024,2024SSRv..220...19N}. This is unsurprising due to their formation through hierarchical mergers of smaller halos, but their rich complexity on an individual basis can significantly change their expected lensing properties. 
This increased complexity provides new and exciting GW lensing phenomenology that can only be accessed through studying lenses rich in structure such as clusters.

Although there have been studies of the impact of small scale structures on both EM \cite{Williams:2023jiq,Venumadhav:2017pps} and GW lensing \cite{Chema_microlensing_2019}, there has been no comprehensive study on the impact that the sub-structure seen in real galaxy cluster models will have on both the rates and observables of lensed gravitational waves.

The paper is structured as follows: Sec.~\ref{sec:lensing} outlines the basics of gravitational lensing formalism, Sec.~\ref{sec:realisticmodel} describes the galaxy cluster models used in this work, as well as our methods of simulating populations of gravitational wave sources in the cluster lensing framework, Sec.~\ref{sec:results} presents the results of this study, Sec.~\ref{sec:implications} discusses the implications of the results, as well as future work, and finally, Sec.~\ref{sec:conclusions} summarizes the findings of this work with some concluding remarks.

\section{Gravitational Lensing}\label{sec:lensing}

Within the geometric optics limit, where the wavelength of the gravitational wave is much smaller than the characteristic size of the lens, GW gravitational lensing phenomenology closely follows that of light \citep{Takahashi_2003}. For a given lensing configuration, the locations of images are determined by the lens equation,
\begin{equation}
    \beta = \theta - \alpha(\theta),
\end{equation}\label{eq:lensequation}
where $\beta$ is the source location, $\theta$ is the image location, and $\alpha(\theta)$ is the deflection angle, determined by the lensing potential $\psi(\theta)$ as $\alpha(\theta) = \nabla \psi(\theta)$. The dimensionless surface mass density, or convergence, is defined as,
\begin{equation}
    \kappa(\theta) = \frac{\Sigma(\theta)}{\Sigma_c},
\end{equation}
where $\Sigma(\theta)$ is the surface mass density of the lens, and $\Sigma_c$ is the critical surface mass density at the redshift of the lens,
\begin{equation}
    \Sigma_c = \frac{c^2 D_S}{4 \pi G D_L D_{LS}},
\end{equation}
where $G$ is Newton's gravitational constant, $c$ is the speed of light, and $D_S$, $D_L$, and $D_{LS}$ are the angular diameter distances to the source, the lens, and between the lens and source respectively. The difference in arrival time between two images $\theta_i$ and $\theta_j$of the same source at a position $\beta$ is
\begin{equation}
    \Delta t_{ij} = \frac{1+z_L}{c} \frac{D_L D_S}{D_{LS}} \big [ t_d(\theta_i, \beta) - t_d(\theta_j, \beta) \big ],
\end{equation}
where $z_L$ is the redshift of the lens, and 
\begin{equation}
    t_d(\theta,\beta) \equiv \bigg[  \frac{(\theta - \beta)^2}{2}  - \psi(\theta) \bigg ]
\end{equation}
is the Fermat potential or time delay surface~\cite{Schneider85, Blandford:1986zz}, and $\psi(\theta$) is the lens potential. 
In this language, the lens equation and image positions $\vec\theta_i$ are simply given by $\left.\partial t_d /\partial \vec\theta \right|_{\vec\theta=\vec\theta_i}=0$. 
Finally, the magnification of a given image is defined as
\begin{equation}
    \mu^{-1} = \left[1-\kappa(\theta)\right]^2 - \gamma(\theta)^2,
\end{equation}
where $\gamma(\theta)$ is the shear. 
This, again, can be derived directly from the time delay surface, i.e., $\mu^{-1}=\mathrm{det}\left(\partial^2t_d/\partial \vec\theta\partial\vec\theta\right)$. 
The points along which $|\mu| \rightarrow \infty$ in the image plane are called \emph{critical curves}, and their equivalent in the source plane are called \emph{caustics}. 
These are seen as the dark regions in Fig.~\ref{fig:muclust}, which show the magnification maps of the three galaxy clusters used in this study (introduced in Sec.~\ref{sec:realisticmodel}) in the source plane (top), and image plane (bottom).

 A quantity that is often used throughout this work is the relative magnification factor, which is simply $\mu_r = |\mu_1 / \mu_2|$, where $\mu_1$ and $\mu_2$ are the brightest and second brightest images of a given source at a location $\beta$. We also define the parity of the images based on the sign of their magnification factors (i.e. a positive parity image has $\mu_i > 0$, whereas a negative parity image has $\mu_i < 0$).

For more simple lenses (such as galaxies), the lensing potential can be described as a singular isothermal sphere (SIS), which has a density distribution that follows
\begin{equation}\label{eq:sisrho}
    \rho_{SIS}(r)  = \frac{\sigma^2_v}{2 \pi G r^{2}},
\end{equation}
where $\sigma_v$ is the velocity dispersion of the dark matter halo. This model will always produce either one image, or when the source is located within the Einstein radius, given by 
\begin{equation}\label{eq:sisrad}
    \theta_{E}  = 4 \pi \frac{\sigma^2_v}{c^2}\frac{D_{LS}}{D_S},
\end{equation}
produces two images of opposite parity. The SIS model is a specific case where the image multiplicity (i.e. the number of images generated based on a single position in the source plane) changes at the location of the Einstein radius. This is not true for most dark matter density profiles \cite{Schneider:1992bmb}, therefore we define the cross section in which a source positions generate more than one image as the \textit{strong lensing cross section} $\sigma_{SL}$. For a given solid angle $\Omega$, we can also define a \textit{source plane optical depth} as 
\begin{equation}
    \tau_{SL} \equiv \sigma_{SL}/\Omega
\end{equation} 
for each lens. 
In the SIS case, this expression is simply given by $\tau_\mathrm{SIS}=\pi \theta_E^2/\Omega$.

\begin{figure}
    \centering
    \includegraphics[width=\linewidth]{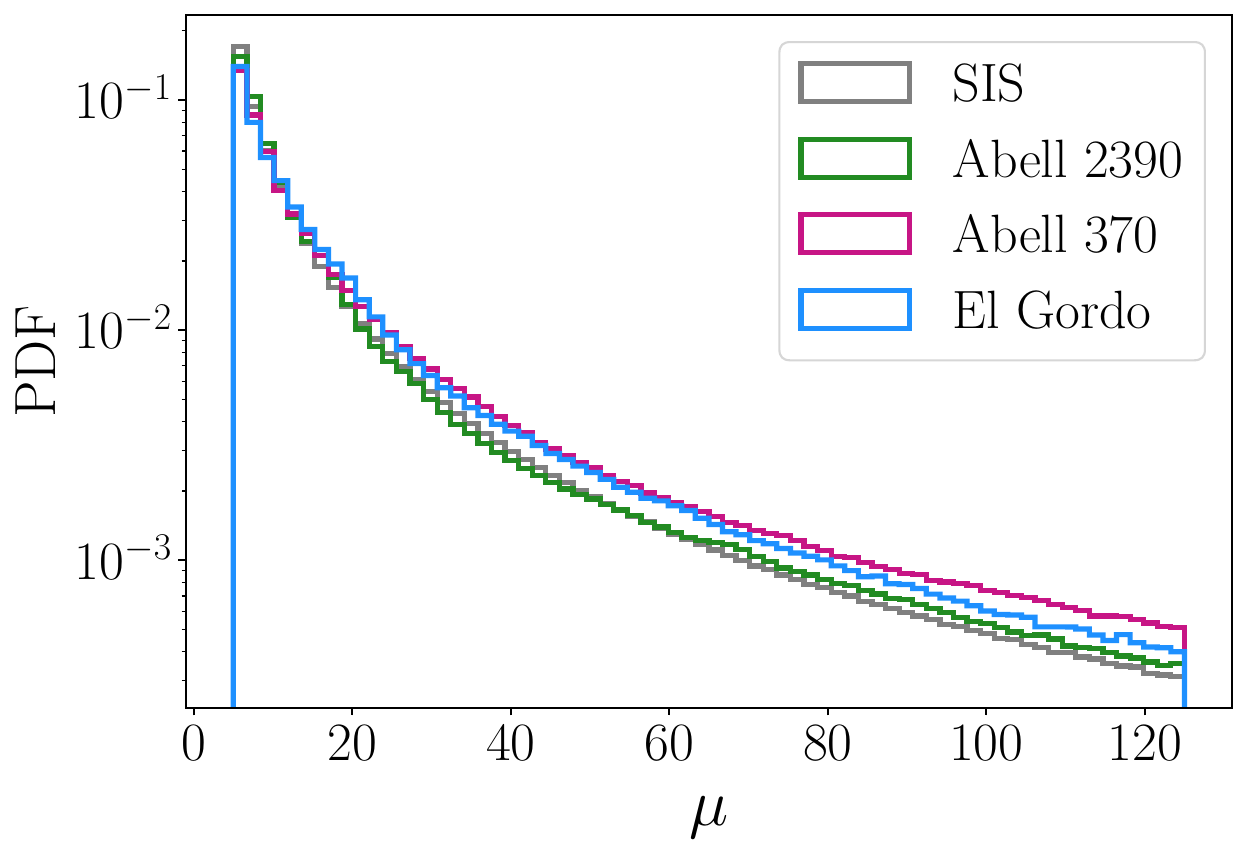}
    \caption{Normalized distribution of image plane magnification factors for a source at $z=3$ for the three clusters in this study, compared against the singular isothermal sphere (SIS) model. 
    The high magnification tail is enhanced in the realistic models.
    }
    \label{fig:mudist}
\end{figure}

\begin{figure*}
    \centering
    \includegraphics[width=\linewidth]{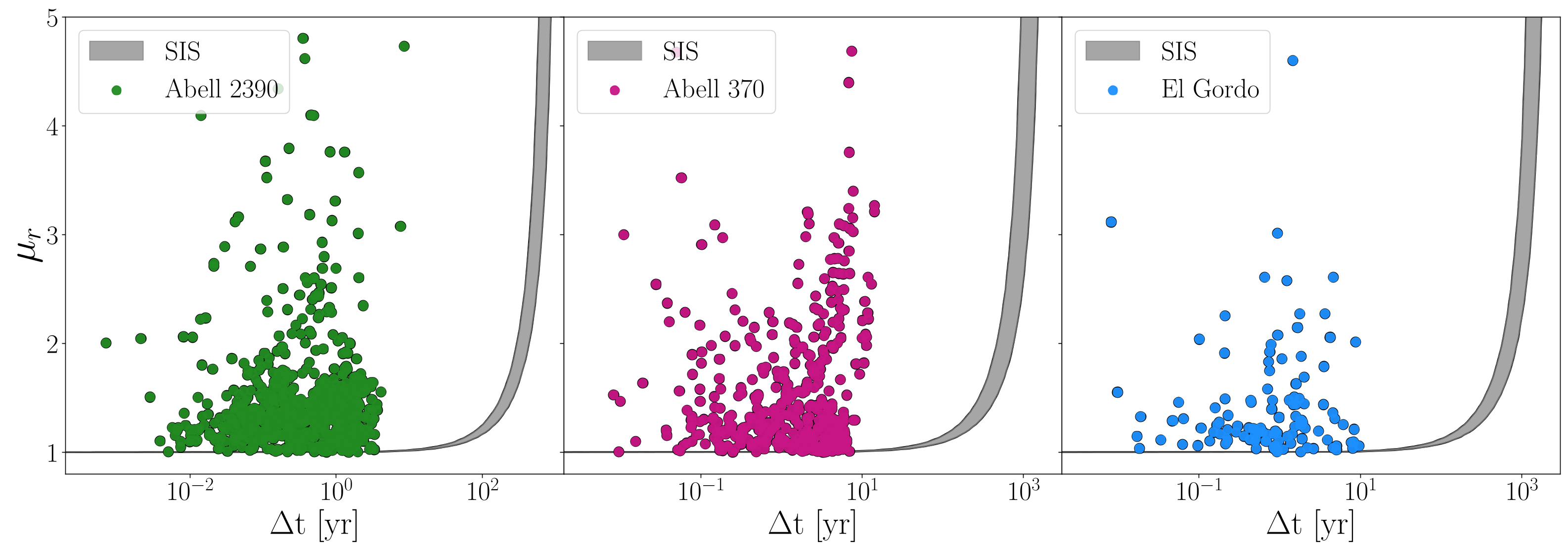}
    \caption{Time delay $t_d$ vs relative magnification factor $\mu_r$ of detectable lensed binary black hole image pairs at O4 sensitivity for Abell 2390 (left), Abell 370 (middle), and El Gordo (right) up to $z=3$. Results are plotted against those from their respective SIS models. Due to the substructure, the maximum time delay of the realistic clusters decreases considerably compared to the isothermal model. The relative magnifications also spread significantly compared to the SIS band, which traces the relation $\mu_r(\Delta t)$ (cf. Eq. \ref{eq:sismur}) over the possible source redshifts.
    }
    \label{fig:tdvsmur_detvssimple}
\end{figure*}

The benefit of using such a model is that it provides a simple analytic description of the desired strong lensing observables for transients, such as time delays, magnification factors, and strong lensing cross sections \cite{Schneider:1992bmb}. This is particularly powerful for calculating the expected rates of detecting strongly lensed sources. 
This is very evident in the case of the SIS model.
Following \cite{Schneider:1992bmb}, our dimensionless impact parameter is defined as $y=\beta/\theta_E$, which leads to the magnification factors for both images 
\begin{equation}\label{eq:sismag}
    \mu_\pm = 1 \pm \frac{1}{y}.    
\end{equation}
The magnification distribution of the SIS model is universal and does not depend on redshift. It corresponds to the gray distribution in Fig. \ref{fig:mudist}. 
On the other hand, the time delay between the two images is given by 
\begin{equation}\label{eq:sisdelt}
    \Delta t = \bigg (4\pi \frac{\sigma_v^2}{c^2} \bigg)^2 \frac{D_{LS} D_L}{cD_S} (1+z_s) 2y  \equiv T_* y\,,
\end{equation}
where $T_*$ is the reference time scale of the lens-source system, and the time delay scales linearly with the impact parameter.
These two analytic expressions for the magnification and time delay in terms of the impact parameter $y$ allow us to write the relative magnification factor as,
\begin{equation}\label{eq:sismur}
    \mu_r = |\mu_+ / \mu_-| = \bigg| \frac{1 + 1/y}{1 - 1/y}           \bigg|  = \bigg|\frac{\Delta t+T_*}{\Delta t-T_*}\bigg|.
\end{equation}
This relation corresponds to gray band in Fig. \ref{fig:tdvsmur_detvssimple}, whose behavior is universal but overall time scale depends on the velocity disperion of the lens.

While modifications to the SIS model can be made to make it a suitably realistic model to describe galaxy lenses (such as including ellipticity and external shear), as we will see in Sec.~\ref{sec:realisticmodel}, it is far too simplistic to accurately model the rich lensing phenomenology seen in most galaxy clusters, which are generally not only non-spherical, but contain many lenses within the same system. Therefore, more sophisticated models are required to match observations of strong lensing in galaxy clusters.

\section{Towards More Realistic Cluster Models}\label{sec:realisticmodel}

To study the impact that galaxy cluster sub-structures have on lensed transient observables, three models were chosen from publicly available cluster lens models  \cite{richard2021}: Abell 2390 to represent a low redshift cluster, Abell 370 to represent a highly concentrated cluster \cite{2007MNRAS.379..190C} at a moderate redshift, and ACT-CL J0102-4915 (also known as 'El Gordo', which will be used in the rest of this work) to represent a high redshift, extended and merging cluster. Whilst no two clusters are the same, we choose these three clusters to represent different extents of structure seen observationally, with higher redshift clusters typically being more rich in structure due to ongoing mergers. Generalizing these results to the global population of clusters remains a difficult task, however there have been studies that employ $N$-body simulations to estimate the cluster lensing optical depths \cite{2020MNRAS.495.3727R}.
These state-of-the-art cluster lens models have been constructed from 
observations with the deep fields of the Hubble Space Telescope and followed-up with the Multi Unit Spectroscopic Explorer survey. 
The magnification maps of these clusters (Fig.~\ref{fig:muclust}) highlight the difference in the structure of these clusters, namely in El Gordo, in which there are two main merging cluster-scale dark matter halos \cite{richard2021}, extended along the South-West to North-East direction. A common feature between all lens models are the perturbations to the cluster-scale potential from galaxy-scale halos, causing intricate deviations from the smooth cluster-scale critical curves and caustics, as well as providing smaller critical curves and caustics in the vicinity of the clusters. The models in this study contain tens to hundreds of member galaxies that are resolved thanks to the tens of spectroscopically resolved multiple image systems \cite{richard2021}. 

\begin{table}[!h]
\caption{Names, redshifts, and total velocity dispersions of the clusters used in this study. }
\label{tab:clusters}
\begin{tabular}{p{1.in} p{1.in} p{1.in}}
\hline
Cluster Name        & $ ~~ z$    & $\sigma_v$ [km s$^{-1}$] \\
\hline
Abell 2390          & $0.228$ & $1882$ \\
Abell 370           & $0.375$ & $1976$  \\ 
El Gordo            & $0.870$ & $1888$ \\
\end{tabular}
\vspace*{-4pt}
\end{table}

These models were created using \texttt{LENSTOOL} \cite{knieb1996, jullo2007, jullo2009}, 
  which employs a parametric modeling approach, which inherently assumes that the lensing potential and mass distribution of the lens can be represented by a sum of parametric density distributions.

The member galaxies are modeled as axially-symmetric dual pseudo-isothermal mass density profiles (dPIE) \cite{eliasdottir2007, suyuhalkola2010}:
\begin{equation}
    \Sigma(R) = \frac{\sigma_{v}^2}{2G} \biggr [ \frac{1}{R}  - \frac{1}{\sqrt{R^2 + r_\mathrm{cut}^2}} \biggr ],
\end{equation}
where R is the radial coordinate, $\sigma_{v}$ is the central velocity dispersion of the galaxy, and $r_\mathrm{cut}$ is the cut radius, at which the slope of the density profile drops steeply. 
The main cluster-scale dark matter halos are described using pseudo-isothermal elliptical mass distributions (PIEMD) \cite{kassiolakovner1993}, which was found to better fit galaxy cluster lenses \cite{grillo2015}:
\begin{equation}
    \Sigma(R) = \frac{\sigma_v^2}{2G} \bigg [ \frac{1}{\sqrt{R(e)^2 + r_\mathrm{core}^2}} \bigg ],
\end{equation}
where $r_\mathrm{core}$ is the core radius, $e$ is the ellipticity, and $R(e)$ is constant over ellipses with ellipticity $e$.

To provide a point of comparison, each cluster model is compared against a singular SIS model (see Eqs.~\ref{eq:sisrho}, \ref{eq:sisrad}, \ref{eq:sismag}, \ref{eq:sisdelt}, \ref{eq:sismur}) to facilitate easy comparisons to the existing GW lensing literature, where this model is used extensively \cite{Ng:2017yiu,Xu:2021bfn,2023lensingO3full}. The velocity dispersion ($\sigma_v$) is taken to be the quadratic sum of all cluster scale mass components in each lens model, which has been shown to closely trace the kinematic velocity dispersion found in cluster lensing systems \cite{richard2021}. Following \cite{eliasdottir2007, richard2021}, the velocity dispersion of cluster scale dPIE dark matter components must be re-scaled by a factor of 1.3 in order to be compared to more general density distributions. The velocity dispersion of a dPIE halo ($\sigma_\mathrm{dPIE}$) is proportional to that of a SIS ($\sigma_\mathrm{SIS}$), 
\begin{equation}
    \sigma_\mathrm{SIS} \approx \sigma_\mathrm{dPIE} \times 1.3 \pm 0.2,
\end{equation}
with an overall, small offset.

\subsection{Strongly Lensed Gravitational Waves}\label{subsec:surveymethod}

With realistic cluster models in hand, we proceed to investigate the impact that realistic dark matter distributions would have on the detection rates of strongly lensed gravitational waves (GW).
Given the small size of binary black holes (BBH) and binary neutron stars (BNS) when compared to the size of the dark matter structures in this work, we consider them to be point sources in our calculation. Similarly, the frequency of a BBH/BNS merger that current ground-based detectors would be sensitive to is significantly smaller than the size of the dark matter halos in this study, meaning that geometric optics will sufficiently model the lensing effects of these signals. In light of this, magnification factors will boost the amplitude of a gravitational wave signal by a factor of $\sqrt{\mu}$. A noteworthy exception to this approximation is the region in which the binary is very close to a caustic, where studies have shown that these highly magnified GWs have not only strong interference and diffraction effects
, leaving observable signatures in current and future observing runs \cite{Bulashenko:2021fes,Lo_highmag,Ezquiaga:2025gkd,Serra:2025kbw}. Whilst there is also the possibility of seeing wave optics phenomena due to compact objects such as stars and black holes near the critical curves of clusters, we do not consider these additional effects in this work.

In order to quantify the differences in strong lensing observables and detection rates between the simple and realistic models, we first start by solving the lens equation for all models at the first redshift in which we get a non-negligible cross section for each model, and up until a redshift of $z=6$, where we found the detection of strongly lensed GWs to be nearly zero for our fiducial population of compact binaries. 
This results in only considering compact binary populations above a minimum redshift of $z=0.5$ for Abell 3290, $z=0.5$ for Abell 370, and $z=1.2$ for El Gordo due to the negligible $\sigma_{SL}$ below these redshifts in the case of realistic lens models, as it is shown in Fig.~\ref{fig:cross_section}. 
Next, we populate the region randomly within the strong lensing cross-section with binaries according to the merger rates, masses and mass ratios 
consistent with the latest LVK catalog, i.e., GWTC-3 
\cite{KAGRA:2021duu}. 
Specifically, this corresponds to a merger history that approximately follows the star formation rate \cite{2014ARA&A..52..415M}, and a mass spectrum parametrized by a (smooth) power law plus peak model within $\sim 5$ and $100M_\odot$ for binary black holes, and a uniform mass distribution for neutron stars. See Appendix \ref{app:pop_cbc} for details. 

Finally, we compute the signal-to-noise ratio (SNR) $\rho$ of each image of the binary assuming current GW detector (O4) sensitivity, (along with the design sensitivity of current detectors (A$+$) and next-generation ground-based facilities such as Einstein Telescope or Cosmic Explorer) \cite{LIGOScientific:2016wof,Hild:2008ng, Hild:2009ns, Hild:2010id} in conjunction with the image properties for the given source location. For the sake of simplicity, we assume a 100 \% detector uptime. Relaxing this assumption would inherently lead to the potential for losing some images that arrive at later times, however we leave quantifying these effects to future studies. 
We quantify a detection of an event if it passes a SNR threshold of $\rho \geq 8$. 
For a given pair of images to be detected, both need to have lensed SNRs of $\rho \geq 8$. 
We also track the relative arrival time of the two brightest images (as well as the rest of them), the parity of all images (i.e. whether their magnification is positive or negative), and the total image multiplicity of the source location.
We neglect the influence of the signal-consistency test value on the detectability of lensed gravitational waves, which will be addressed in future investigations \cite{Chan:2024qmb}.

\begin{table*}[!t]
\caption{Detection rates per year of gravitational waves behind our three fiducial clusters, Abell 2390, Abell 370 and El Gordo and up to $z=6$. We consider different source populations, binary black holes (BBH) and binary neutron stars (BNS), and different detector sensitivities from present sensitivities (O4, now), to design sensitivities of current detectors (A$+$, expected 2027), to next-generation ground-based detectors (XG, expected 2035+).}%
\label{tab:rates}
\begin{tabular}{p{0.7in} | w{c}{0.7in} | w{c}{0.7in} | w{c}{0.7in}  w{c}{0.7in} | w{c}{0.7in}  | w{c}{0.7in}| w{c}{0.7in}}
\hline
Cluster   & Model     & BBH, O4  & BNS, O4 & BBH, A$+$  & BNS, A$+$ & BBH, XG  & BNS, XG\\
\hline
\rowcolor{lightgray} \textbf{Abell 2390}  &  Realistic         & $7.8 \times 10^{-7}$ & $2.8\times 10^{-9}$ & $3.8 \times 10^{-6}$ & $4.7\times 10^{-8}$ & $1.6 \times 10^{-5}$ & $2.2\times 10^{-5}$\\
 & SIS                                                         & $6.9 \times 10^{-5}$ & $8.0\times 10^{-6}$& $1.9 \times 10^{-4}$ & $3.2\times 10^{-5}$& $2.2 \times 10^{-3}$ & $1.4\times 10^{-3}$ \\   
\hline
\rowcolor{lightgray} \textbf{Abell 370}   & Realistic        & $3.1\times 10^{-6}$ & $1.9\times 10^{-7}$ & $8.5 \times 10^{-6}$ & $7.8\times 10^{-7}$& $4.0 \times 10^{-5}$ & $6.0\times 10^{-5}$ \\ 
 & SIS                                                       & $5.8 \times 10^{-5}$ & $6.6\times 10^{-6}$ & $1.7 \times 10^{-4}$ & $2.4\times 10^{-5}$& $2.0 \times 10^{-3}$ & $1.2\times 10^{-3}$ \\ 
\hline
\rowcolor{lightgray} \textbf{El Gordo}  & Realistic          & $7.9 \times 10^{-7}$ & $1.2\times 10^{-8}$& $2.9 \times 10^{-6}$ & $1.1\times 10^{-7}$& $2.4 \times 10^{-5}$ & $2.34\times 10^{-5}$ \\
 & SIS                                                       & $1.3 \times 10^{-5}$ & $1.5\times 10^{-6}$ & $3.9 \times 10^{-5}$ & $4.6\times 10^{-6}$& $5.8 \times 10^{-4}$ & $2.9\times 10^{-4}$\\
\end{tabular}
\end{table*}

\section{\label{sec:results}Results}

After introducing the three clusters under investigation in this study and the lensing formalism, we proceed to present our main results.

\subsection{Relative Magnifications and Time Delays}\label{subsec:td_mur}

In the context of time domain transients, the main observables of a multiply imaged system is the relative magnifications and time delays of the different image pairs. 
For flux-limited surveys, the most relevant observables are those of the two brightest images. 
We present the relative magnifications and time delays for the two brightest images of transients occurring behind our three study-case clusters in Fig.~\ref{fig:tdvsmur_detvssimple}. 
We compare those results with those obtained with a simple SIS model (grey color band). Although only the detected image pairs are presented in Fig.~\ref{fig:tdvsmur_detvssimple}, the entire population of image pairs compared against the SIS model is presented in the Appendices (Fig.~\ref{fig:tdvsmur_detvsundet}).
This figure shows two distinct features between the realistic and simple models: the mean time delay of the two brightest images of the realistic models is drastically lower than those of the simple model, and there is a new population of short time delay, high relative magnification images in the realistic model. These are all to be expected given that the total halo mass from the simple model is now broken up into the smaller halos found in the realistic model. These small halos are typically responsible for producing short time delay images with high relative magnification factors, even though the total mass of the cluster remains unchanged. 
If one were to try and infer the lens mass producing such an image pair using Eq.~\ref{eq:sismur}, the result could be misinterpreted as coming from an isolated galaxy-scale lens.
It should be noted that although cluster-scale halos are capable of producing short time delay images, this only occurs when sources are near caustics, implying high magnification factors with $\mu_r\sim1$ (seen in the short $\Delta t$ tail of the clusters in Fig.~\ref{fig:tdvsmur_detvssimple}), meaning that the cluster scale components cannot be responsible for the high $\mu_r$, short $\Delta t$ images in Fig.~\ref{fig:tdvsmur_detvssimple}.

\subsection{Redshift Dependent Cross Sections}\label{subsec:cross_sec}

\begin{figure}
    \centering
    \includegraphics[width=\linewidth]{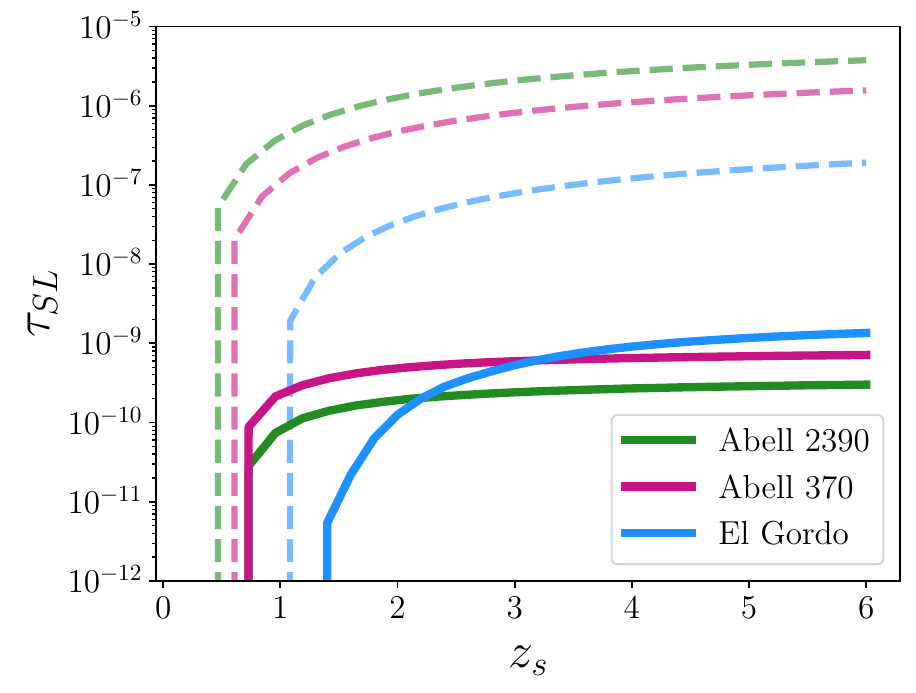}
    \caption{Optical depth ($\tau_{SL}$) of the three cluster models (solid lines) compared against the SIS predictions (dashed lines).
    }
    \label{fig:cross_section}
\end{figure}

There are two competing effects that in combination, provide a non-trivial impact on the detectability of strongly lensed sources behind clusters: the loss of strong lensing cross section through having a non-singular distribution of lens components (Fig.~\ref{fig:cross_section}), as well as the lengthening of critical curves produced by the same effect, resulting in an increase in the high magnification tail of the magnification distribution (Fig.~\ref{fig:mudist}). In a simple case, the strong lensing cross section of a halo that either has any non-negligible ellipticity, or a cored profile such as a PIEMD (or Navarro-Frenk-White (NFW)) profile \cite{navarro1997,Navarro:1996gj}), there is both a drop in cross section, as well as an increase in the lensing efficiency \cite{wrightbrainerd1999}. A noteworthy consideration is that along with having smaller cross sections for a fixed lensing configuration (i.e. a fixed $D_{s}, D_{L}$ and $D_{LS}$), cored profiles are significantly less efficient at image splitting \cite{wrightbrainerd1999}. Therefore, they only produce non-negligible strong lensing cross-sections at higher values of $D_{LS}/D_{L}$ than SIS (or even Singular Isothermal Ellipsoids (SIE) lenses \cite{kormann1994}), thus capturing noticeably smaller amounts of potential sources along a fixed cluster line of sight. These discrepancies are evident upon comparing the number of detected strongly lensed images and image pairs (Tab.~\ref{tab:rates}), where we see that despite the increase in lensing efficiency, the loss in cross section seems to hinder the detection rates behind realistic cluster models.

Another key feature to note is the trend of the clusters with more substructure producing larger discrepancies from the simple model. This is yet again explained by the total mass of the realistic models being distributed throughout more separated halos. 
For example, the majority of the mass in Abell 2390 is contained within the main central dark matter halo (left panels in Fig.~\ref{fig:muclust}), where as El Gordo's is distributed throughout many merging dark matter halos (right panels in Fig.~\ref{fig:muclust}) that, given its high redshift, are still in the process of merging, and are thus separated further from each other than those of the other two clusters in this study. 
Simply put, this means that relaxed, virialized clusters will be better represented by simple lens models, which introduces a redshift dependence to the deviation of the lensed observables of realistic cluster models from their simple counterparts, which is expected given the known redshift dependence on galaxy cluster structure from both observations and $N$-body simulations.

\subsection{Detection Rates}\label{subsec:rates}

\begin{figure}
    \centering
    \includegraphics[width=\linewidth]{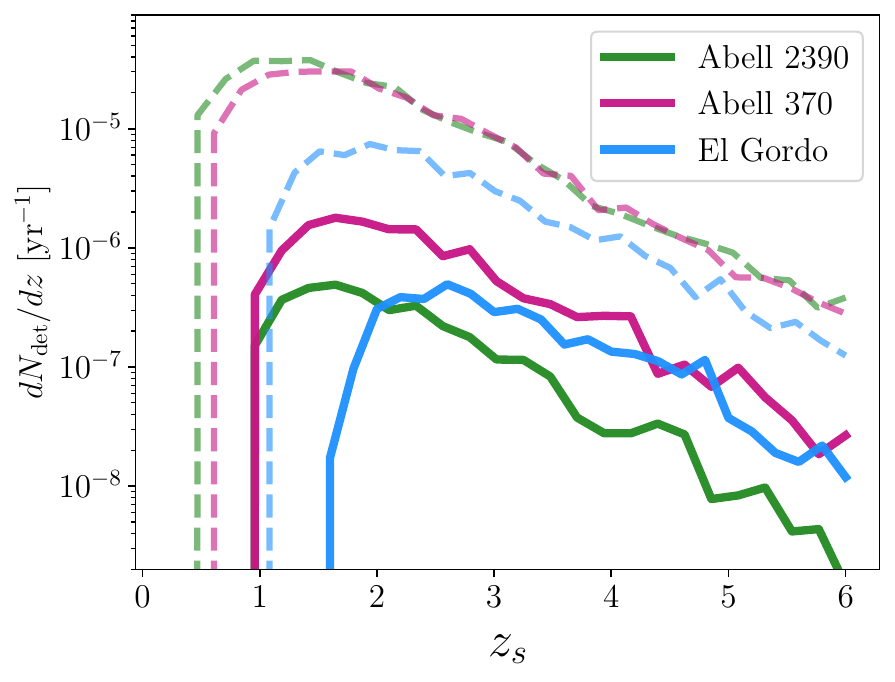}
    \caption{Differential gravitational wave detection rates as a function of redshift for BBH sources behind 
    the realistic models (solid lines), and SIS models (dashed lines). 
    Detection rates are computed for current detector's sensitivity (O4).
    }
    \label{fig:rates}
\end{figure}

The rates of detecting both a single image, as well as pairs of strongly lensed images were found to be consistently higher in the case of SIS lenses than the realistic models. The number of detected lensed events are summarized in Tab.~\ref{tab:rates} for sources up to $z=6$, for different detector sensitivities (O4, A$+$ , and XG). 
The differential detection rate presented in Fig.~\ref{fig:rates} shows that the peak of the distribution of detected sources is at a higher redshift for the realistic models as opposed to the SIS. This is most likely due to the realistic models being inefficient at image splitting for sources close to the lens, thus having very small regions of high magnification at low source redshifts. The overall drop in detection in the case of the realistic models is most likely due to the significantly smaller strong lensing cross section when compared to the SIS. Interestingly, the weak lensing or single-image regime of the SIS model detectable images is slightly lower, but comparable to the number of detectable strongly lensed BBH image pairs, whereas the realistic models vary by up to an order of magnitude from the strongly lensed images (see details in Tab.~\ref{tab:rates_weak} in App.~\ref{app:single_image}). We also find negligible numbers of BNS detections at O4 and A+ sensitivities in the single image regime for both models. The detection horizon in the single image regime increases from approximately $z_s \sim 1.7$ in the SIS case to $z_s \sim 4$ in the realistic case (again, see details in Fig.~\ref{fig:rates_weak} in the appendix) due to increase in high magnification optical depth in this regime.

We find that the realistic models also produce a higher fraction of detectable positive parity images than their negative parity counterparts (30-40\% more than the predicted ratio of 1 for the SIS case). This is in agreement with studies of the impact of substructure on EM observables \cite{Williams:2023jiq} , which also finds an increase in the fraction of positive vs. negative party images as the amount of substructure increases.

\subsection{Magnification Distribution}\label{subsec:magdist}

We find that the cases due to substructure inevitably increasing the length of critical curves, and thus the opportunity for high magnification factors, there is an increase in the high magnification tail of the normalized magnification distribution (Fig.~\ref{fig:mudist}). Unsurprisingly, this trend seems to depend on the complexity of the lens in question, which can easily be seen in Fig.~\ref{fig:mudist}, which from left to right, shows increases every time we go to a slightly more complex lens.

Furthermore, we find that the tail of both distributions still approximately follow the theoretical predictions of a power-law-like behavior of $N(\mu) \propto \mu^{-2}$ \cite{richard2021, Blandford:1986zz}. A noteworthy caveat is that the amount of pixels that we find with high magnification factors is heavily dependent on the resolution of the lens model. This is mainly due to a good deal of the caustic lengthening being done by smaller lenses such as member galaxies, meaning that it is inherently harder to resolve their high magnification regions. This effect will be especially apparent in lenses with large amounts of visible substructure (such as El Gordo).

\subsection{Image multiplicity}\label{subsec:multiplicity}

\begin{figure}
    \centering
    \includegraphics[width=\linewidth]{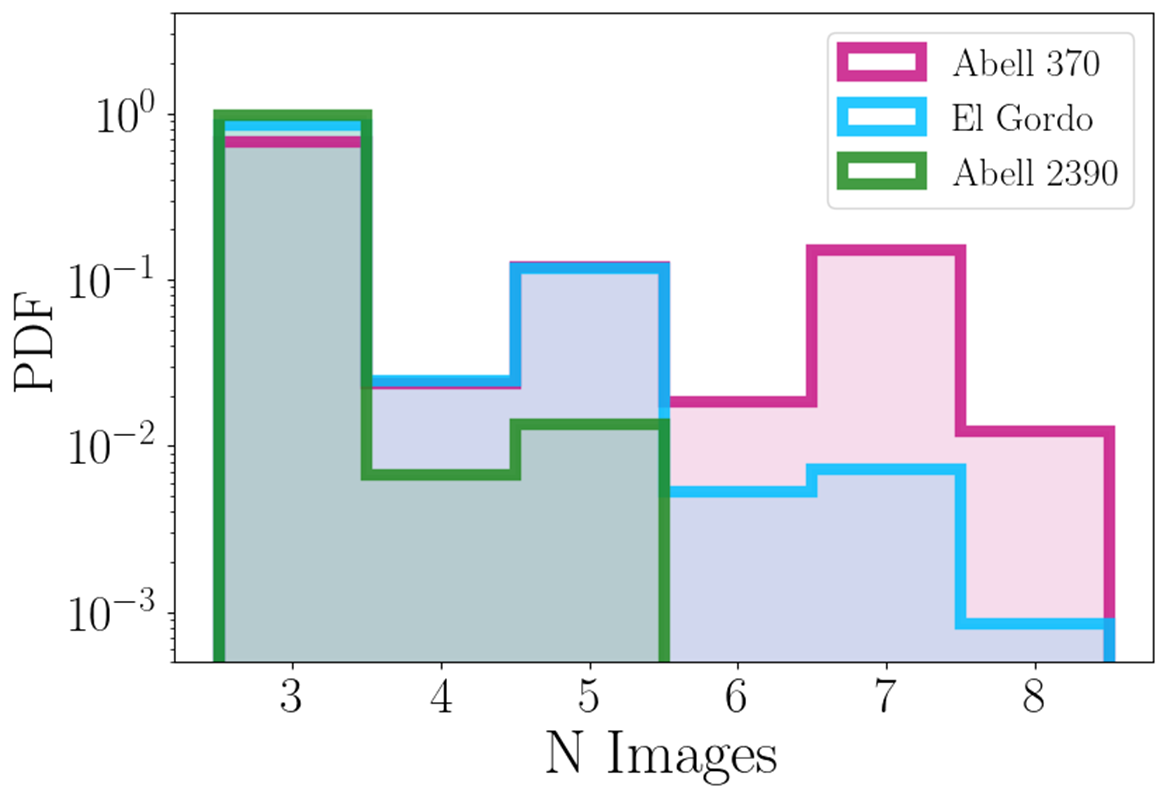}
    \caption{Normalized distribution of image multiplicity regions in the source plane for Abell 2390 (green), Abell 370 (pink), and El Gordo (blue). Note that the SIS model can only produce one or two images. 
    }
    \label{fig:allmult}
\end{figure}

One of the largest difference between the SIS models and the cluster models used in this work are the image multiplicities found within each source plane. As mentioned in Sec.~\ref{sec:lensing}, the SIS model only produces two images. However, even by simply considering more realistic halos such as a PIEMD model, single halos can produce either 1, 3, or 5 images. When considering many halos, the overlapping strong lensing cross sections can produce even higher image multiplicities (Fig.~\ref{fig:allmult},~\ref{fig:a370mult}).

Additionally, the realistic models allow for significantly higher magnification factors in the single image regime, as opposed to the $\mu<2$ limit for any source at $y>1$ for the SIS case. This can be clearly seen outside of the cusps of the source plane magnification distributions of Fig.~\ref{fig:muclust}, and extend far beyond the caustics.

The magnification distribution of the different image multiplicity regions in the source plane is shown in the Appendices in Fig.~\ref{fig:multmu}. Across all models, the highest multiplicity regions (in this case, being the 7 and 8 image regions), tend to have smaller areas of high magnification. This has important implications for sources that must be highly magnified in order to be seen at cosmological distances, such as binary neutron star mergers (BNS). It has been predicted that the first detected strongly lensed BNS will have a magnification factor of at least $\mu\gtrsim100$ \cite{Smith:2022vbp}. This would mean that the higher image multiplicity regions would be statistically disfavored for highly magnified images, which could help inform EM follow-up campaigns designed to detect the appearance of future images.

\section{\label{sec:implications}Implications}

The results presented in the last section have far reaching implications for many science cases associated to lensed transients: from the inference of the lens model from the data, to the prospects of observing lensed transients with future facilities. 
We discuss them in different subsections.

\subsection{Lens Model Reconstruction}\label{subsec:reconstruction}

Although there are hopes of reconstructing galaxy-scale lens models using the detection of strongly lensed GWs (ideally supplemented with EM observations \cite{Hannuksela:2020xor}), the results of our study disfavor the prospect of succeeding in this endeavor for cluster-scale lenses. The main deterrent is the presence of both cluster and galaxy-scale dark matter halos (Fig.~\ref{fig:muclust}), resulting in strong degeneracies in the $\Delta t$ and $\mu_r$ distribution that one would need to assume to reconstruct the lens mass strictly from GWs (Fig.~\ref{fig:tdvsmur_detvssimple}), especially if only two images are detected.

However, the distribution of time-delays that a lensing configuration can produce for its two brightest images has a very sharp edge at high time-delays (Fig.~\ref{fig:tdvsmur_detvssimple}). This means that although as previously discussed, short time-delay images with non-unity relative magnification factors can be caused by small lenses embedded in larger lenses, bright images with long time delays can only be produced by massive lenses. 

An observation of such an image pair could provide a lower bound on the mass of the lens, which would be invaluable for rapid EM followups by allowing for the preferential targeting of high-mass known lenses \cite{lenscat2024} or structures. A hindrance to such a detection is the time delay between the two images, which can be upwards of $\sim 10$ years, thus requiring continuous observations (ideally from multiple GW detectors) to capture the signal from the second image, and identify the image pair as coming from the same source. These increased time delays between images also increase the false alarm probability for sources lensed by clusters \cite{Wierda:2021upe}.

\subsection{Moderately magnified single-image regions} 

As an interesting aftermath of the intricate structure of realistic cluster lenses, the single-image region around the multiple image caustic has a sizable area with moderately high magnification, $\mu\gtrsim2$. This can be seen directly in the shaded region around the main caustics of each cluster in Fig. \ref{fig:muclust}. 
Also, as we have shown in the results section, the rates associated with this region are comparable to the multiple-image analog (see details in App. \ref{app:single_image}). 
Moreover, the realistic models produce larger magnifications that significantly increase the detection horizon compared to the idealized model. 
Having a single lensed image posses new challenges for the identification of the event as lensed, as most transient search methods rely on finding multiple instances of the original event.
In the particular case of GWs, an unnoticed lensed event will be interpreted as being closer and having a larger mass,  causing apparent outliers in the given population of binaries \cite{LIGOScientific:2021izm}. This will bias the true source properties, affecting the inferred modeling of the astrophysical population.
A possible aid in identifying this moderately magnified single-image cases would be to cross-correlate with cluster catalogs. For example, \texttt{lenscat} allows for an easy cross-match with known cluster scale lenses~\cite{lenscat2024}.

\subsection{Effects of Subhalos}\label{subsec:subhalos}

An effect that is unexplored in this work is that of dark matter subhalos. It has been long predicted that massive dark matter halos are comprised of smaller subhalos. The density of these subhalos typically follow an NFW profile for typical cold dark matter models \cite{Navarro:1996gj}. but could form other interesting density distributions in self interacting, warm, fuzzy, or wave dark matter models \cite{Viel_2013,Tulin_2018,Niemeyer_2020,hui2021}. These subhalos could potentially be very important for the prospects of detecting strongly lensed transients due to their proposed existence near critical curves, thus potentially lengthening them and creating higher chances of seeing a very highly magnified transient \cite{Williams:2023jiq, Venumadhav:2017pps}. This effect of critical curve lengthening could also increase the chances of seeing diffraction effects in transients of longer wavelengths such as gravitational waves, which could provide smoking gun evidence for lensing, even with a single detected image. The dark matter model also plays a role in setting the lower bounds on subhalo masses (with models such as self interacting and warm dark matter suppressing power on small scales). Therefore, the detection of wave optics effects (in which the wavelength of the signal is comparable to the size of the lens) would provide an invaluable probe on the mass range of subhalos, allowing for the constraint of current dark matter models and mass ranges.
Exploring these effects outside of the scope of this paper due to the small scale of the subhalos, which might produce wave optics effects that are not captured in these simulations \cite{Villarrubia-Rojo:2024xcj}, and is therefore left to future work.

\subsection{Extensions Towards Effects on Global Rates}\label{subsec:globalrates}

Although it is possible to calculate the deviations between detection rates from simple models and realistic cluster models on an individual case, constraints on the global detection rate still remain elusive. The main limitation is the lack of lens models of clusters both across cosmic time, and across their entire mass range. 

Past studies employing N-body simulations find that, on average, the strong lensing cross sections of galaxy clusters is reasonably well described by an SIS halo when the SIS is normalized to have the same mass enclosed as the mass within the virial radius of a simulated cluster \cite{2020MNRAS.495.3727R}. 
However, we find that prescribing an SIS halo of the same measured velocity dispersion as realistic models provides an inadequate description of the strong lensing observables. Given that the optical depth of high mass clusters ($M \sim 10^{15} M_\odot$) such as the clusters used in this work is low when compared to the contributions from galaxies or groups of galaxies ($M\sim10^{11}$--$10^{13}M_\odot$) \cite{2020MNRAS.495.3727R}, our findings are unlikely to impact significantly the total rate of transients that are strongly lensed by the full halo-mass function,
or change the total strong lensing optical depth from similar studies. 
It would be interesting to consider in the future how our results extend to lighter clusters in the $\sim10^{14}M_\odot$ mass range as those have a bigger contribution to the optical depth, especially at low redshifts.

As mentioned in Sec.~\ref{subsec:cross_sec}, the separation of cluster-scale halos contributes substantially to the loss of strong lensing cross section. Given the fact that higher redshift clusters are more likely to have separated cluster halos due to them still being in the process of merging \cite{2024SSRv..220...19N}, there is a non-trivial correlation between the redshift of the cluster and its strong lensing cross section. The loss of cross section through main halo separation outweighs the boost in the probability of having highly-magnified images through the lengthening of critical curves and caustics, thus reducing the rate of detected strongly lensed transients. 

Despite the present lack of large catalogs of publicly detailed cluster models, the upcoming James Webb Space Telescope (\emph{JWST}) program ``\emph{JWST} Cluster SLICE'' (GO 5594 PI: Mahler) aims to target this problem specifically by targeting 182 clusters across a large redshift range ($0.2\leq z \leq 1.9$) and mass range. Although the proposed depth of the survey will be $AB \sim 27$ mag., this should be sufficient to detect enough multiple image pairs to constrain the large-scale morphology of the clusters across the relatively large sample.

It should be stressed that as important as it is to have high-quality lens models, the public availability of the complete lens model is imperative to facilitate studies such as these.

\section{Summary and Conclusions}\label{sec:conclusions}

Gravitational lensing is a natural tool to explore further into the distant Universe and discover new transients. It provides insight into their origin, with the sources themselves acting as probes of the cosmic history. 
Because of the difficulties in accounting for the impact of the complexities of cluster structure on the global populations of GW lensing observables, the focus of most studies has been on galaxy-scale lenses.
However, groups and cluster of galaxies hold a large fraction of the mass in the Universe, can act as very efficient lenses. Examples of this are their use as ``gravitational telescopes'' to observe the furthest galaxies \cite{2023ApJ...957L..34W, 2024arXiv241113640K}. 
The difficulty is still that those systems exhibit a complex morphology that can effect the lensing observables.
We present the impact of using real galaxy cluster lens models as foreground lenses as opposed to the simple, single halos that have been used extensively in the transient lensing literature. Three clusters were selected: Abell 2390, Abell 370, and El Gordo, and were compared against singular isothermal sphere (SIS) models of the same total velocity dispersion. 
Our cluster lensing study is general for any point source, although we particularize the detectability discussion for gravitational waves.
The key results are as follows:

 \begin{itemize}

     \item The strong lensing cross section of each cluster lens model is significantly smaller than its SIS counterpart.

     \item The added structure of the realistic lens models increases the high-magnification tail of the magnification distribution, thus increasing the amount of highly-magnified images. It also produces a higher image multiplicity.

     \item Despite the increase of increased lensing efficiency, the detection rate of gravitational waves behind each realistic cluster model is approximately an order of magnitude lower than its SIS counterpart in the cases of both strongly lensed primary images and image pairs. 
     
     \item The single image regime of the SIS models are found to have slightly lower but comparable detection rates to the number of detected strongly lensed BBH image pairs. The number of detected BBH single images varies by up to an order of magnitude from the number of detected BBH image pairs in the realistic models. This is due to the interplay of the larger area but also the fact that single-image regions can reach large magnifications in the realistic case. There are negligible number of BNS detections from the single image regime at O4 and A+ sensitivities.

     \item The realistic models produce more detectable positive parity images, increasing the expected ratio of positive to negative parity images from unity in the SIS model to $\sim$ 30-40 \% more positive parity images than negative.

     \item The distribution of time delays and relative magnifications of the realistic models spreads over a much wider range than the narrow SIS prediction. Moreover, the maximum time delay decreases significantly, complicating distinguishing these lenses from isolated galaxies.

  \end{itemize}

This study presents a step forward towards understanding the rich phenomenology of lensing by clusters. 
Our results show that clusters are efficient lenses to expand the detection horizon of transient surveys. 
Their possibly high image multiplicity and long time delays serve as unique fingerprints for the identification of lensed transients, although we also find that with just a pair of images it might be difficult to discriminate from lensing by galaxies. 
The non-negligible high magnification single-image regions also points for cross-correlating transient and cluster catalogs. 
It is important to note that although the rates of gravitational waves have been explored in this study, these results should apply more generally to any compact time domain transients whose merger history roughly follows the star formation rate such as fast radio bursts and gamma ray bursts.

\begin{acknowledgments}
The authors would like to thank Johan Richard for both providing the lens models used in this work, as well as providing helpful insights into using the models for our applications. The authors would also like to thank Lukas Furtak, Srashti Goyal, Daniel Holz, Guillaume Mahler, Anupreeta More, Andrew Robertson, Liliya Williams, and Miguel Zumalacárregui for their helpful comments and suggestions.
The Center of Gravity is a Center of Excellence funded by the Danish National Research Foundation under grant No. 184.
This project was supported by the research grant no. VIL37766 and no. VIL53101 from Villum Fonden, and the DNRF Chair program grant no. DNRF162 by the Danish National Research Foundation.
This project has received funding from the European Union's Horizon 2020 research and innovation programme under the Marie Sklodowska-Curie grant agreement No 101131233. 
J.M.E. is also supported by the Marie Sklodowska-Curie grant agreement No.~847523 INTERACTIONS. 
The Tycho supercomputer hosted at the SCIENCE HPC center at the University of Copenhagen was used for supporting this work.
\end{acknowledgments}

\appendix

\section{Population of compact binaries}
\label{app:pop_cbc}

In order to compute the lensing rates behind a cluster we need to make some assumptions about the population of compact binaries. 
We inform our fiducial models by the data from the latest GW catalog, GWTC-3 \cite{KAGRA:2021duu}. 
This data constraints the merger rate history of binary black holes only below $z\sim1$ and that of binary neutron stars only below $z\sim0.1$. 
For simplicity we assume that both populations have a merger rate history $R(z)$ that follows the star formation rate:
\begin{equation}
    R(z)= R_0\cdot C_0 \frac{(1+z)^\alpha}{1+\left(\frac{1+z}{1+z_p}\right)^{\alpha+\beta}}\,,
\end{equation}
which is described by a rising slope $\alpha$, a peak at $z_p$ and a decaying slope $\beta$. In parallel with the Madau \& Dickson parametrization \cite{Madau:2014bja}, we choose $\alpha=2.7$, $\beta=5.6$ and $z_p=1.9$. $C_0$ is a function of $\alpha$, $\beta$ and $z_p$ which ensures that $R(0)=R_0$. 
Following GWTC-3 we fix the comoving local merger rate of binary black holes to $30 \mathrm{yr}^{-1}\mathrm{Gpc}^{-3}$ and that of binary neutron stars to $100 \mathrm{yr}^{-1}\mathrm{Gpc}^{-3}$.

In terms of the mass function of binary black holes, we assume that they follow a smooth, power law plus peak model. The specific values we choose are $\alpha=-3.4$, $\beta_q = 1.1$, $m_{min}=8.75$, $m_{max}=150$, $\lambda_{peak}=1.4\times10^{-8}$, $\mu_m=34$, $\sigma_m=3.6$, $\delta_m=2$.

We parametrize the binary neutron star mass function with a simple uniform distribution between $1M_\odot$ and $2.2M_\odot$. We also choose a uniform distribution in mass ratios.

\section{Image multiplicity maps and magnification distributions}
\label{app:multiplicities}

\begin{figure*}
    \centering
    \includegraphics[width=\linewidth]{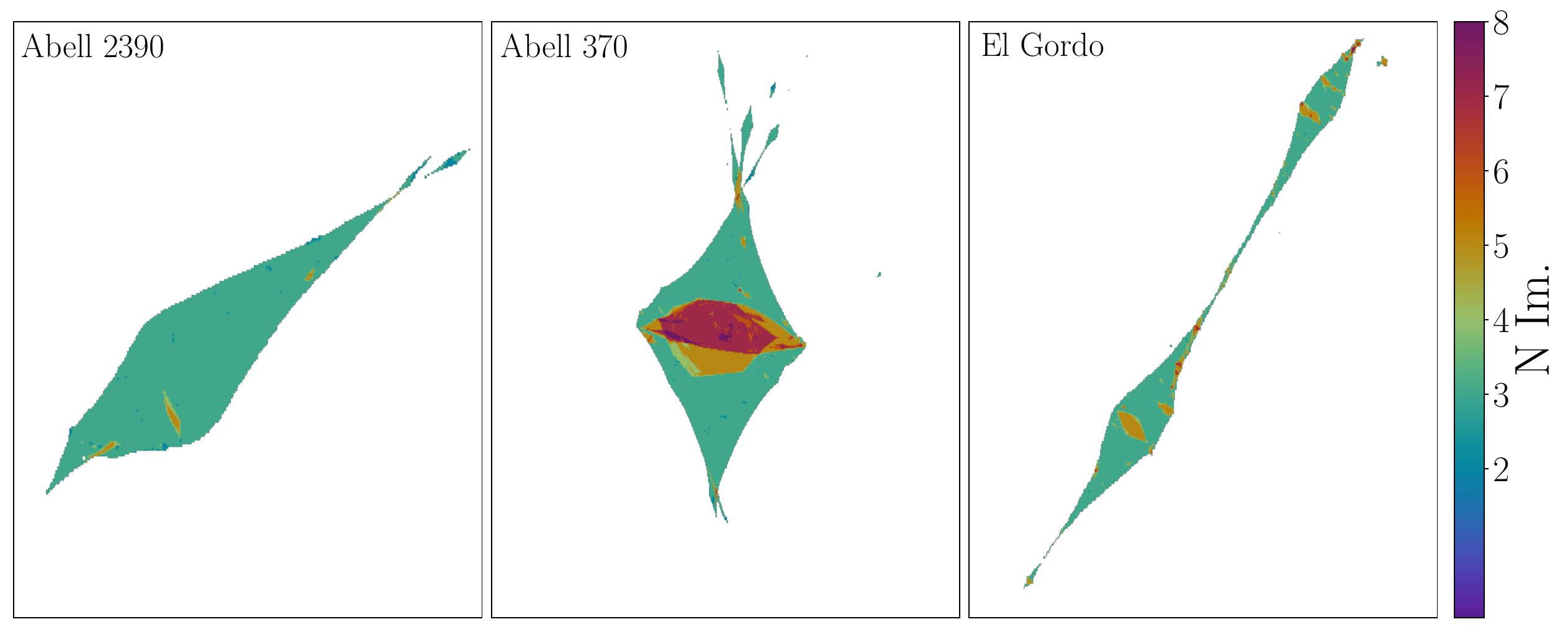}
    \caption{Image multiplicity maps of Abell 2390 (left), Abell 370 (middle), and El Gordo (right) for a source at $z=3$.
    }
    \label{fig:a370mult}
\end{figure*}

The number of images is a key lensing observable. To complement the results presented in the main text, here we plot the image multiplicity maps for the three clusters in Fig. \ref{fig:a370mult}. We find that whereas Abell 2390 and El Gordo have relatively small higher image multiplicity regions, a large portion of the central region of the strong lensing region Abell 370 is dominated by higher image multiplicity regions. A noteworthy detail is that whereas in this study, we find that Abell 2390 is only capable of producing a maximum of 5 images, both Abell 370 and El Gordo are capable of generating up to 8 images, but only do so in very small fractions of the source plane (c.f. Fig.~\ref{fig:a370mult}). 

\begin{figure*}
    \centering
    \includegraphics[width=\linewidth]{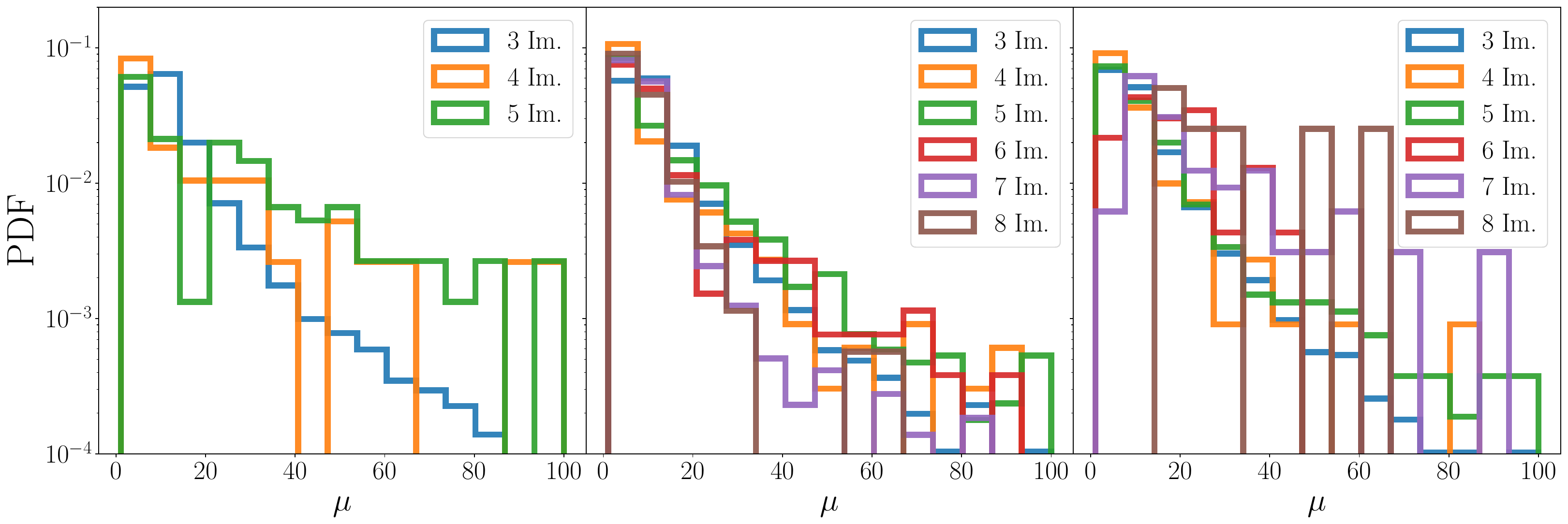}
    \caption{Normalized distribution of the magnifications of the brightest image in each multiplicity region in the source plane for Abell 2390 (left), Abell 370 (middle), and El Gordo (right) for sources at $z=3$. Note that the SIS model can only produce one or two images. 
    }
    \label{fig:multmu}
\end{figure*}

We also present in Fig. \ref{fig:multmu} the magnification distribution for the different image multiplicities.

\subsection{Single Image Regime}
\label{app:single_image}

\begin{figure}[!h]
    \centering
    \includegraphics[width=\linewidth]{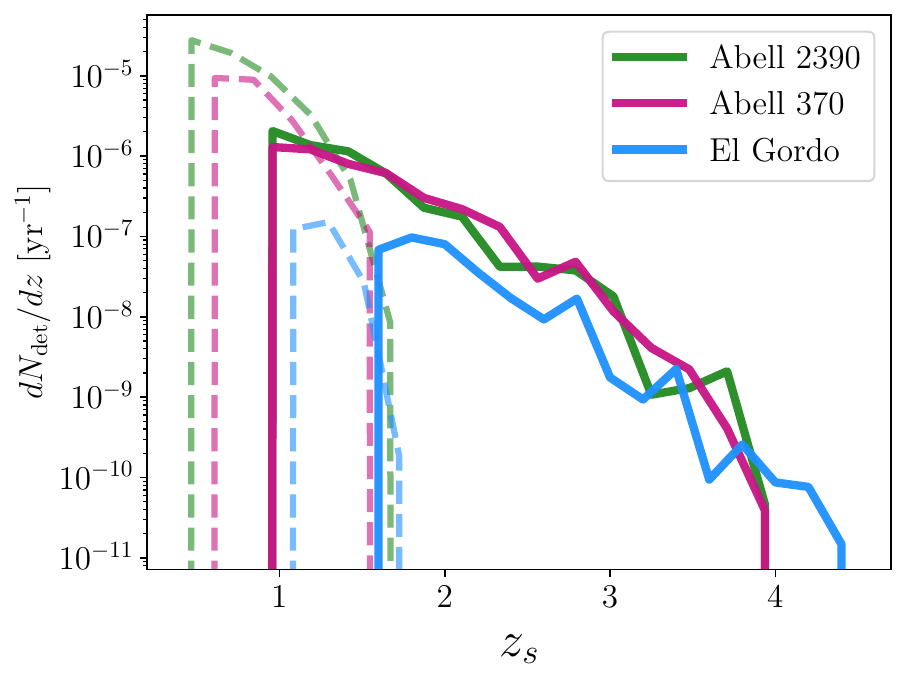}
    \caption{Like Fig.~\ref{fig:rates}, but for the single image regime. 
    Detection rates are computed for current detector's sensitivity (O4).
    }
    \label{fig:rates_weak}
\end{figure}

While the main results of this work pertain to the observability of strongly lensed transients, an interesting discrepancy between the SIS and realistic models is the single image regime, more commonly referred to as the ``weak lensing regime". Whilst the SIS model can only produce maximum magnification factors of 2 in the single image regime, the top row of Fig.~\ref{fig:muclust} shows that in realistic models, you can get magnification factors much higher than this. This will have a dramatic impact on the detection horizon for singly imaged sources, allowing realistic models to probe higher redshifts than the SIS model as seen in Fig.~\ref{fig:rates_weak}, with the total detection rates presented in Tab.~\ref{tab:rates_weak}.

\begin{table*}[!t]
\caption{Like Tab.~\ref{tab:rates}, but only considering the single image regions behind the clusters in this study.}%
\label{tab:rates_weak}
\begin{tabular}{p{0.7in} | w{c}{0.7in} | w{c}{0.7in} | w{c}{0.7in}  w{c}{0.7in} | w{c}{0.7in}  | w{c}{0.7in}| w{c}{0.7in}}
\hline
Cluster   & Model     & BBH, O4  & BNS, O4 & BBH, A$+$   & BNS, A$+$  & BBH, XG  & BNS, XG\\
\hline
\rowcolor{lightgray} \textbf{Abell 2390}  &  Realistic        & $1.3 \times 10^{-6}$ & $ < 10^{-10}$ & $1.6 \times 10^{-5}$ & $< 10^{-10}$ & $7.4 \times 10^{-4}$ & $1.4\times 10^{-4}$\\
 & SIS                                                        & $1.5 \times 10^{-5}$ & $< 10^{-10}$& $1.8 \times 10^{-4}$ & $< 10^{-10}$& $2.3 \times 10^{-2}$ & $1.5\times 10^{-3}$ \\   
\hline
\rowcolor{lightgray} \textbf{Abell 370}   & Realistic        & $1.1\times 10^{-6}$ & $< 10^{-10}$ & $1.1 \times 10^{-5}$ & $< 10^{-10}$& $4.8 \times 10^{-4}$ & $1.0\times 10^{-4}$ \\ 
 & SIS                                                       & $5.8 \times 10^{-5}$ & $< 10^{-10}$ & $1.2 \times 10^{-4}$ & $< 10^{-10}$& $2.1 \times 10^{-2}$ & $8.6\times 10^{-4}$ \\ 
\hline
\rowcolor{lightgray} \textbf{El Gordo}  & Realistic          & $6.6 \times 10^{-8}$ & $< 10^{-10}$& $1.7 \times 10^{-6}$ & $< 10^{-10}$& $1.5 \times 10^{-4}$ & $1.5\times 10^{-5}$ \\
 & SIS                                                       & $6.3 \times 10^{-8}$ & $< 10^{-10}$ & $8.7 \times 10^{-6}$ & $< 10^{-10}$& $5.7 \times 10^{-3}$ & $4.9\times 10^{-5}$\\
\end{tabular}
\end{table*}

\section{Selection biases in the distribution of time delays and relative magnifications}

\begin{figure*}
    \centering
    \includegraphics[width=\linewidth]{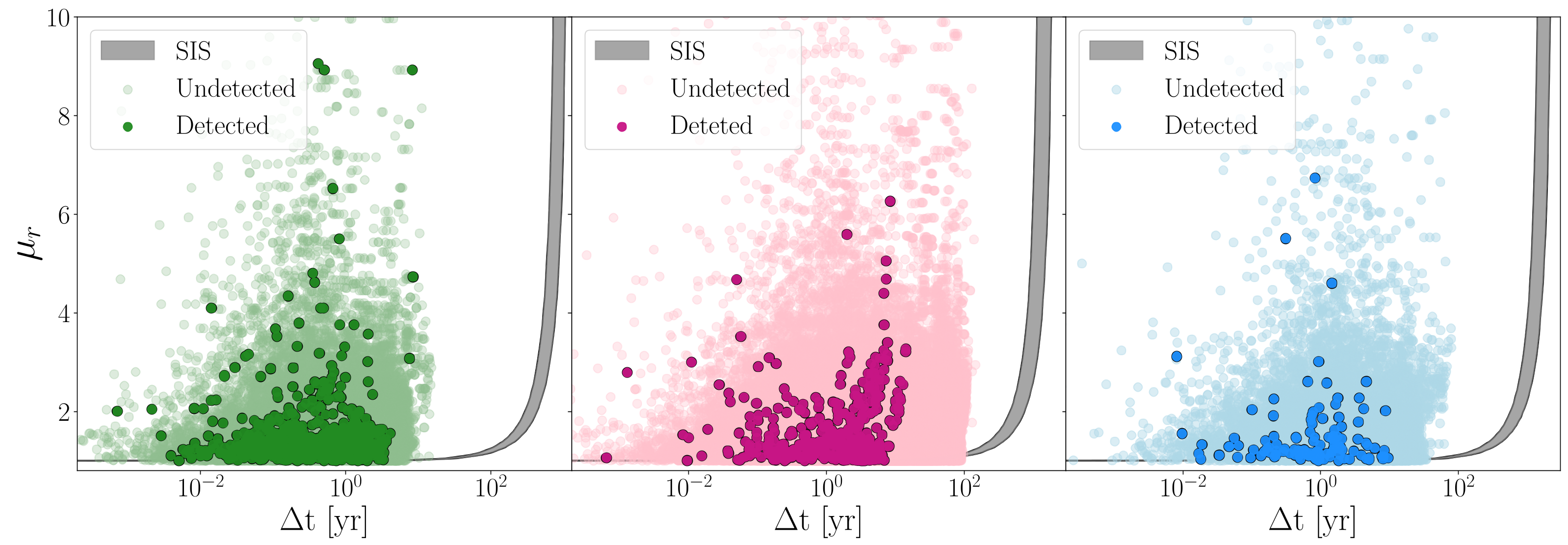}
    \caption{Absolute value of time delay $\Delta t$ vs relative magnification factor $\mu_r$ for Abell 2390 (left), Abell 370 (middle), and El Gordo (right) of both detected and undetected brightest image pairs.
    }
    \label{fig:tdvsmur_detvsundet}
\end{figure*}

The observed population of lensed transients might be different from the intrinsic population of lensed sources. In the case of GWs, current detectors are only sensitive to relatively nearby unlensed systems ($z\lesssim1$). Therefore, lensed pairs with higher magnification and shorter time delays are preferred. 
This can be seen in Fig. \ref{fig:tdvsmur_detvsundet}, where we overlay in gray the lensed population with the detected population displayed in Fig. \ref{fig:tdvsmur_detvssimple}. 
The entire lensed distribution has a larger spread and reaches higher time delays. Nonetheless, the maximum time delay is still lower than the prediction from the single singular isothermal sphere model.

\bibliography{references}

\end{document}

%% file: new_commands.tex
\newcommand{\ba}{\begin{eqnarray}}
\newcommand{\ea}{\end{eqnarray}}
\newcommand{\be}{\begin{equation}}
\newcommand{\ee}{\end{equation}}

\definecolor{grey}{rgb}{0.4,0.4,0.4}
\definecolor{dullmagenta}{rgb}{0.4,0,0.4}
\definecolor{darkblue}{rgb}{0,0,0.4}
\definecolor{midblue}{rgb}{0,0,0.5}
\definecolor{midred}{rgb}{0.5,0,0}
\definecolor{orange}{rgb}{1,0.5,0}
\definecolor{lightbrown}{rgb}{0.75,0.5,0.25}
\definecolor{tan}{cmyk}{0.14,0.42,0.56,0}
\definecolor{djunglegreen}{cmyk}{0.99,0,0.52,0}
\definecolor{lightgreen}{rgb}{0,1,0}
\definecolor{olivegreen}{cmyk}{0.64,0,0.95,0.40}
\definecolor{midgreen}{rgb}{0.0,0.675,0.0}
\definecolor{darkgreen}{rgb}{0,0.5,0}

\definecolor{grey}{rgb}{0.4,0.4,0.4}
\definecolor{dullmagenta}{rgb}{0.4,0,0.4}
\definecolor{darkblue}{rgb}{0,0,0.4}
\definecolor{midblue}{rgb}{0,0,0.5}
\definecolor{midred}{rgb}{0.5,0,0}
\definecolor{orange}{rgb}{1,0.5,0}
\definecolor{lightbrown}{rgb}{0.75,0.5,0.25}
\definecolor{tan}{cmyk}{0.14,0.42,0.56,0}
\definecolor{djunglegreen}{cmyk}{0.99,0,0.52,0}
\definecolor{lightgreen}{rgb}{0,1,0}
\definecolor{olivegreen}{cmyk}{0.64,0,0.95,0.40}
\definecolor{midgreen}{rgb}{0.0,0.675,0.0}
\definecolor{darkgreen}{rgb}{0,0.5,0}
\definecolor{ultramarine}{rgb}{0.07, 0.04, 0.56}
\definecolor{cadmiumgreen}{rgb}{0.0, 0.42, 0.24}
\definecolor{indigo(dye)}{rgb}{0.0, 0.25, 0.42}
\definecolor{bostonuniversityred}{rgb}{0.8, 0.0, 0.0}
\definecolor{burgundy}{rgb}{0.5, 0.0, 0.13}
\definecolor{carnelian}{rgb}{0.7, 0.11, 0.11}